\documentclass[a4paper,11pt]{article}
\usepackage{geometry}
\usepackage[svgnames]{xcolor}
\usepackage[colorlinks=true,urlcolor=black,linkcolor=blue,citecolor=blue,bookmarks=false]{hyperref}
\usepackage{amsfonts,amsmath,amssymb}
\usepackage{setspace}
\usepackage{graphicx}
\usepackage{dsfont}
\usepackage{graphicx}
\usepackage{bm}
\usepackage[font=small,labelfont=bf]{caption}
\renewcommand*{\thepage}{\footnotesize\arabic{page}}
\usepackage{color}
\usepackage{enumitem}
\usepackage{boldline}
\usepackage{changepage}
\usepackage{cite}
\usepackage{textcomp}
\usepackage[title]{appendix}
\usepackage{url}
\usepackage[OT1]{fontenc}
 \usepackage{mathpazo}
\usepackage{authblk}
\usepackage{breakurl}
\usepackage{fancyhdr}
\usepackage[export]{adjustbox}
\usepackage[percent]{overpic}
\usepackage{multirow}
\usepackage{colortbl}
\usepackage{diagbox}

\definecolor{lightgray}{rgb}{0.92 0.92 0.92}
\definecolor{darkgray}{rgb}{0.8 0.8 0.8}

\title{\bf Response functions as a new concept to study local dynamics in traffic networks}

\author{Shanshan Wang \thanks{shanshan.wang@uni-due.de}, Michael Schreckenberg and Thomas Guhr}
\affil{\textit{Faculty of Physics, University of Duisburg--Essen, Lotharstra\ss e 1, 47048 Duisburg, Germany}}

\date{\today}
 
\begin{document}
\maketitle

\noindent {\bf Abstract.}
Vehicle velocities in neighbouring road sections are correlated with memory effects. We explore the response of the velocities in the sequence of sections to a congestion in a given section and its dynamic characteristics. To this end, we transfer the concept of response functions from previous applications in finance to traffic systems. The dynamical characteristics are of particular interest. We identify two phases, a phase of transient response and a phase of long-term response. The transient response is pronounced when considering the backward propagation of heavy congestions but almost vanishes for forward propagation. For each response phase, we find a linear relation between the velocity response and the congestion correlator, implying that the correlation of congestion is most likely the cause for the velocity response. We also construct a susceptible-decelerated-withdrawing model mathematically inspired by the susceptible-infectious-recovered (SIR) model in epidemiology to describe the transient response. We find that the heavy congestion on a section propagates forward and backward at a similar rate, but the forward sections are more likely to recover from the effect of heavy congestion than the backward sections.

\vspace{0.5cm}

\vspace{1cm}


\noindent\rule{\textwidth}{1pt}
\vspace*{-1cm}
{\setlength{\parskip}{0pt plus 1pt} \tableofcontents}
\noindent\rule{\textwidth}{1pt}

\section{Introduction}
\label{sec1}

A complex system, composed of mutually interacting agents or constituents, typically shows correlation on different scales between the measured time series which allow one to infer structured dynamical information. A traffic network may be viewed as such a complex system, the role of the constituents is played by the road sections where velocities and flows are measured, and the interactions are induced by the moving vehicles. We recently put forward a variety of studies~\cite{Wang2020,Wang2021,Gartzke2022,Wang2022} in this spirit and identified dominant and subdominant collectivities in time and space. A new challenge, not occurring in, for example, financial networks, is that the informations analyzed for the virtual network of correlated road sections must be mapped back onto the real topology of the road map, i.e. there are two networks in traffic. In finance, response functions have been introduced with considerable success to analyze the reaction of the system to specific events as well as the corresponding time scales~\cite{Bouchaud2003,Wang2016a,Wang2016b,Wang2017,Benzaquen2017,Henao2021}. It is the purpose of this contribution to transfer the concept of response functions to traffic science. In finance, the response functions helped to identify the time scales of the weak non-Markovian effects. In traffic, those effects are obviously strong and are, as we will show, clearly revealed by the response functions.  Events for which the traffic system's response is of particular interest, are congestions~\cite{Krause2017,Zhang2019,Tang2018}, accidents~\cite{Saladie2020}, road construction~\cite{Fei2016}, and presence or absence of trucks~\cite{Han2015}. Such events might be described by an indicator function, i.e. by zero or one if the event is active or passive, respectively. Then, the correlations of these indicators also become informative observables.

The response function not only reveals the sensitivity or vulnerability of the response variable, but also in turn reflects the influence of the triggering event. In a traffic system, the congestion, especially heavy congestion or just a traffic jam, as the typical event affects the velocity of vehicles. This effect will be propagated to a road section different from the congestion site by the motion of vehicles~\cite{Bellocchi2020,Saberi2020,Saeedmanesh2017}. The propagation on the space takes time and differentiates the directions of traffic flows, leading to a spatiotemporal response of velocity to congestion. Here we focus on heavy congestions and the response variable is the vehicle velocity on a motorway section. The response function measures the average velocity change due to the effect of heavy congestion. By empirically analyzing the traffic data, we aim to reveal the basic characteristics of velocity responses among neighbouring sections at first hand. Furthermore, we put forward a susceptible-decelerated-withdrawing (SDW) model to capture and mathematically describe the analyzed response functions. Clearly, this mathematical description does not provide a microscopic explanation of the effects. Nevertheless, it captures the shape of the empirical curves very well and thus facilitates a precise characterization by the model parameters. The mathematics of our SDW model is inspired by that of the susceptible-infectious-recovered (SIR) model in epidemiology, but we point out that there is no direct correspondence of the variables. 

This paper is organized as follows. In section~\ref{sec2}, we describe the traffic data used in this study. In section~\ref{sec3}, we introduce the concept of response functions, empirical analyze two kinds of behavior in velocity responses with respect to different propagation directions of heavy congestion, and discuss the possible cause to induce the velocity response. In section~\ref{sec4}, we construct our SDW model to describe the transient response and reveal the reason why the backward and forward propagation of heavy congestion leads to the different behavior in velocity responses.

\section{Data description}
\label{sec2}

\begin{figure}[tb]
\begin{center}
\includegraphics[width=0.6\textwidth]{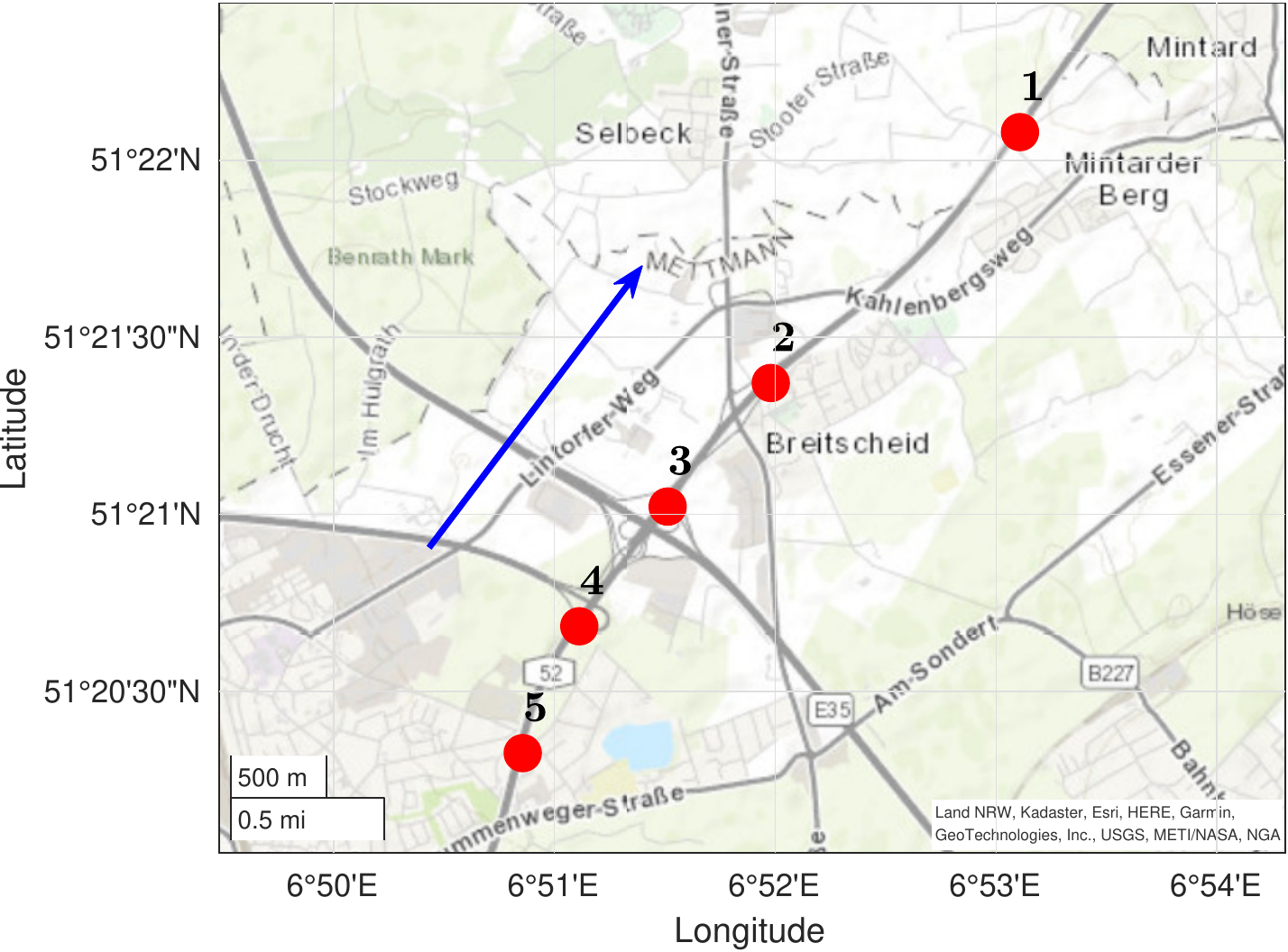}
\caption{Motorways near Breitscheid (NRW). Five selected motorway sections are numbered 1,2,$\cdots$,5. The direction of traffic flow is from section 5 to section 1 as the blue arrow indicates. The base map is provided by Esri~\cite{esri} and the complete map is developed by Matlab~\cite{matlab}.}
\label{fig1}
\end{center}
\vspace*{-0.5cm}
\end{figure}

The traffic data used in this study, including the flow and velocity of vehicles, were measured by inductive loop detectors on each lane of each section on the motorway network of North Rhine-Westphalia (NRW). The flow $q_{i,l}(t)$ records the number of vehicles passing one lane $l$ of one section $i$ during one-minute interval $t$. The corresponding velocity $v_{i,l}(t)$ is averaged over the $t$-th one-minute interval. In view of the data quality influenced by missing values, we use the traffic data from 64 discontinuous workdays in 2017 and restrict the data to the range between 00:00 and 23:59 of each considered day. We select 5 motorway sections near Breitscheid in NRW (shown in figure~\ref{fig1}) for exploration.

Considering each motorway section as an individual, we aggregate the traffic flow across multiple lanes 
\begin{equation}
q_i(t)=\sum_l q_{i,l}(t)  \ .
\label{eq21}
\end{equation}
into the total flow of a section. For each lane, the traffic flow divided by the corresponding velocity gives rise to the flow density, which quantifies the number of vehicles in a unit distance. The summation of flow densities across multiple lanes yields the total number of vehicles in a unit distance , i.e., the flow density for a section,
\begin{equation}
\rho_{i}(t)=\sum_l \frac{q_{i,l}(t)}{v_{i,l}(t)} \ .
\label{eq22}
\end{equation}
This finally results in the velocity of a section as
\begin{equation}
v_i(t)=\frac{q_i(t)}{\rho_{i}(t)} \ .
\label{eq23}
\end{equation}
The same value of traffic flow may correspond to both the free and the congested traffic phases, as demonstrated by the flow-density diagram in Kerner's three-phase traffic theory~\cite{Kerner2012}. In contrast, a value of velocity corresponds either the congested or the free traffic phase and therefore distinguishes the two traffic phases well, illustrated by the velocity-density diagram~\cite{Kerner2012}. In view of this, in the following, we study the response of velocity $v_i(t)$ to the congested traffic phase.

\section{Velocity response to heavy congestion}
\label{sec3}

To the best of our knowledge, the response function is a new concept for studying traffic system. Therefore, we first introduce this function in section~\ref{sec31}, and present the dynamic characteristics of velocity response among neighbouring motorway sections by empirically analyzing the traffic data in section~\ref{sec32}. We separate the response phase in order to find the possible causes for the velocity responses in section~\ref{sec33}. We further explore the duration of heavy congestion that mainly contributes to the congestion correlators.

\subsection{Response functions}
\label{sec31}

On a motorway in a free traffic phase, vehicle velocities reach 80 km/h or more. We define a heavy congestion by setting a threshold of 10 km/h which is roughly comparable to the human walking velocity~\cite{walking}. Hence we introduce the indicator function
\begin{equation}
\varepsilon_{j}(t)=\left\{
\begin{array}{rl}
1,&\mathrm{if~}v_j(t)<10\mathrm{~km/h,}\\
0,&\mathrm{otherwise\ ,}
\end{array}
\right. 
\label{eq31}
\end{equation}
for section $j$, which converts a time series of velocities to a time series of indicators, where $\varepsilon_{j}(t)=1$ represents the heavy congestion present on the motorway section $j$ at time $t$. For two different motorway sections $i$ and $j$, the correlation of indicators with time lag $\tau$, i.e. the congestion correlator, reads
\begin{equation}
\Theta_{ij}(\tau)=\frac{1}{T-\tau}\sum\limits_{t=1}^{T-\tau}\tilde{\varepsilon}_i(t+\tau)\tilde{\varepsilon}_j(t) \ ,
\label{eq32}
\end{equation}
where $\tilde{\varepsilon}_j(t)$ represents the indicator $\varepsilon_{j}(t)$ normalized to zero mean and unit standard deviation. 

Conditioned on the heavy congestion on a motorway section $j$ at time $t$, the response function gives, on average, the velocity change on another motorway section $i$ at a later time $t+\tau$,
\begin{equation}
R_{ij}(\tau)=\frac{\sum\limits_{t=1}^{T-\tau}\big(v_i(t+\tau)-v_i(t)\big)\varepsilon_j(t)}{\sum\limits_{t=1}^{T-\tau}\varepsilon_j(t)} \ ,
\label{eq33}
\end{equation}
extending the definition in finance~\cite{Bouchaud2003,Wang2016a,Wang2016b,Wang2017,Grimm2019,Henao2021} to traffic systems. We notice that the response function is an average over the times $t$ at which a congestion occurs as measured by the indicator $\varepsilon_j(t)$. It is useful to normalize by the total number of these events.

\begin{figure}[tb]
\begin{center}
\includegraphics[width=1\textwidth]{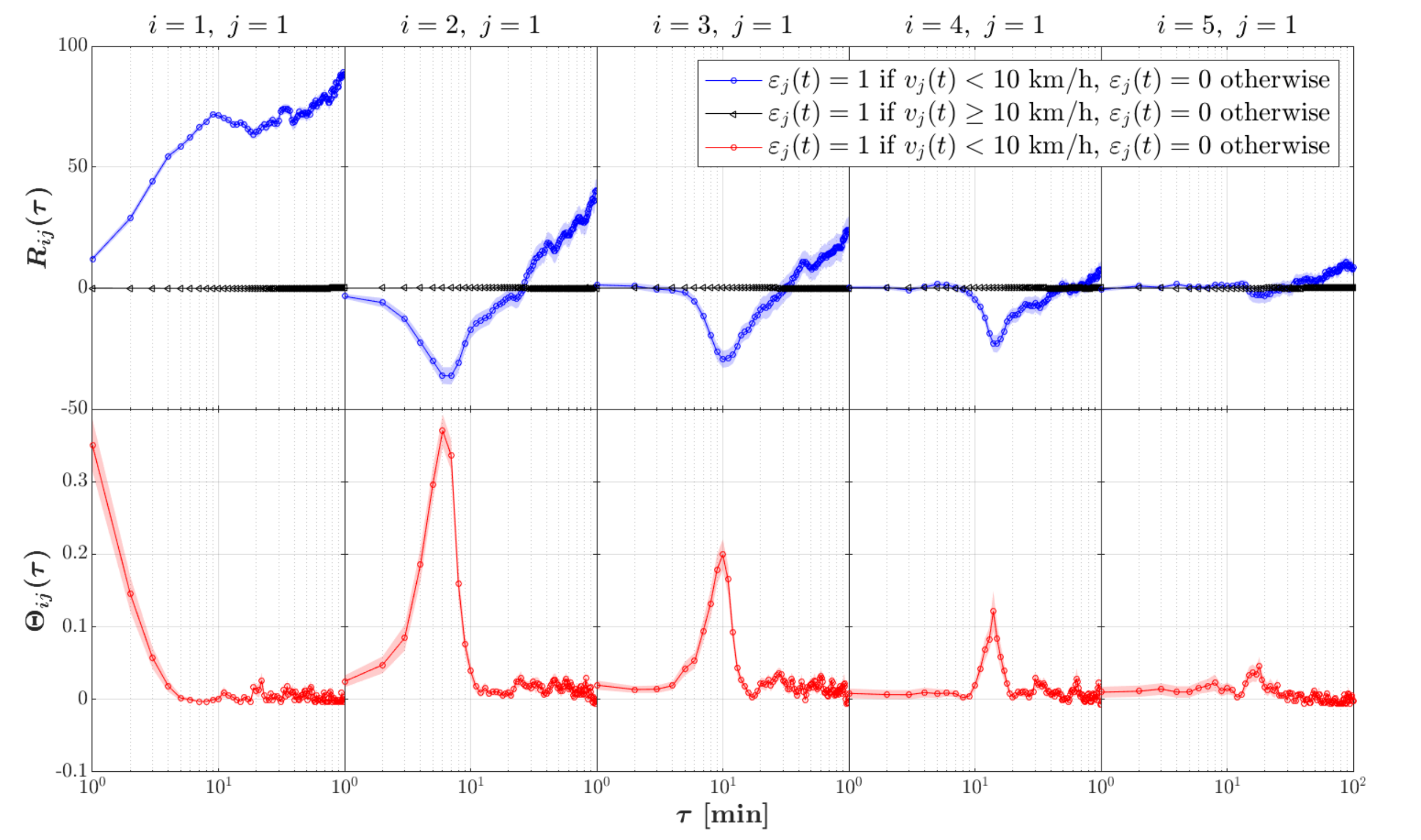}
\caption{The dependences of the velocity response $R_{ij}(\tau)$ and of the congestion correlator $\Theta_{ij}(\tau)$ on time lag $\tau$, where $i=1,~\cdots,~5$ and $j=1$. The light blue and light red shadow area, respectively, indicates the standard errors of the responses and of the correlators.}
\label{fig2}
\end{center}
\vspace*{-0.5cm}
\end{figure}

As an example, figure~\ref{fig2} shows how the velocities of five sections respond to the heavy congestion of section 1. The response of section 1 to itself, i.e. the self-response, only shows positive values with time lag $\tau$. Here we restrict the maximal time lag to 300 min, i.e., 5 hours. Quite differently, the responses to other four neighbouring sections, i.e. cross-responses, present negative values at beginning and then increase to positive values with $\tau$. For different positions of responding sections, the characteristics of velocity responses vary significantly, implying a dependence of responses on the space. Thus, we focus on the response on spatially subsequent sections. 

In finance, we~\cite{Wang2016a,Wang2016b} referred to $R_{ij}(\tau)$ for $i=j$ as self-response and for $i\neq j$ as cross-response. In the present context this terminology is not meaningful as we consider a spatial sequence of sections. Similarly, we used the terminology self- and cross-correlators for $\Theta_{ij}(\tau)$ with $i=j$ and $i\neq j$, respectively. Here the self-correlator $\Theta_{ii}(\tau)$ reveals the time-lagged correlation of heavy congestion on the same section, which decays with time as shown in figure~\ref{fig2}. When it first reaches zero in the interval between 5 to 10 minutes, the correlation of heavy congestion disappears, hence a congestion-based stop-and-go traffic wave (i.e., a sequence of different moving traffic jams) or the initial narrow moving jam~\cite{Kerner2012} on this section disappears. With time increasing, the vehicles in front of this section escape from and the vehicles behind this section enter into the heavy congestion, leading to the next stop-and-go traffic wave or next narrow moving jam. Consequently, small fluctuations show up in the curve of the self-correlator after it first reaches zero. Meanwhile, successive several narrow moving jams are very likely to transform into a wide moving jam~\cite{Kerner2012}. On the other hand, the time-lagged correlation of heavy congestion on spatially subsequent sections, indicated by the cross-correlator $\Theta_{ij}(\tau)$ with $i\neq j$, exhibits the abrupt peaks at the beginning of $\tau$ in figure~\ref{fig2}. The height of peaks, similar to the case of velocity responses, depends on the position of a section relative to its correlated section.  Since we focus on spatially subsequent sections, the congestion correlator in the following means the cross-correlator except for a specific indication.

\subsection{Two kinds of response behavior}
\label{sec32}

\begin{figure}[tb]
\begin{center}
\includegraphics[width=1\textwidth]{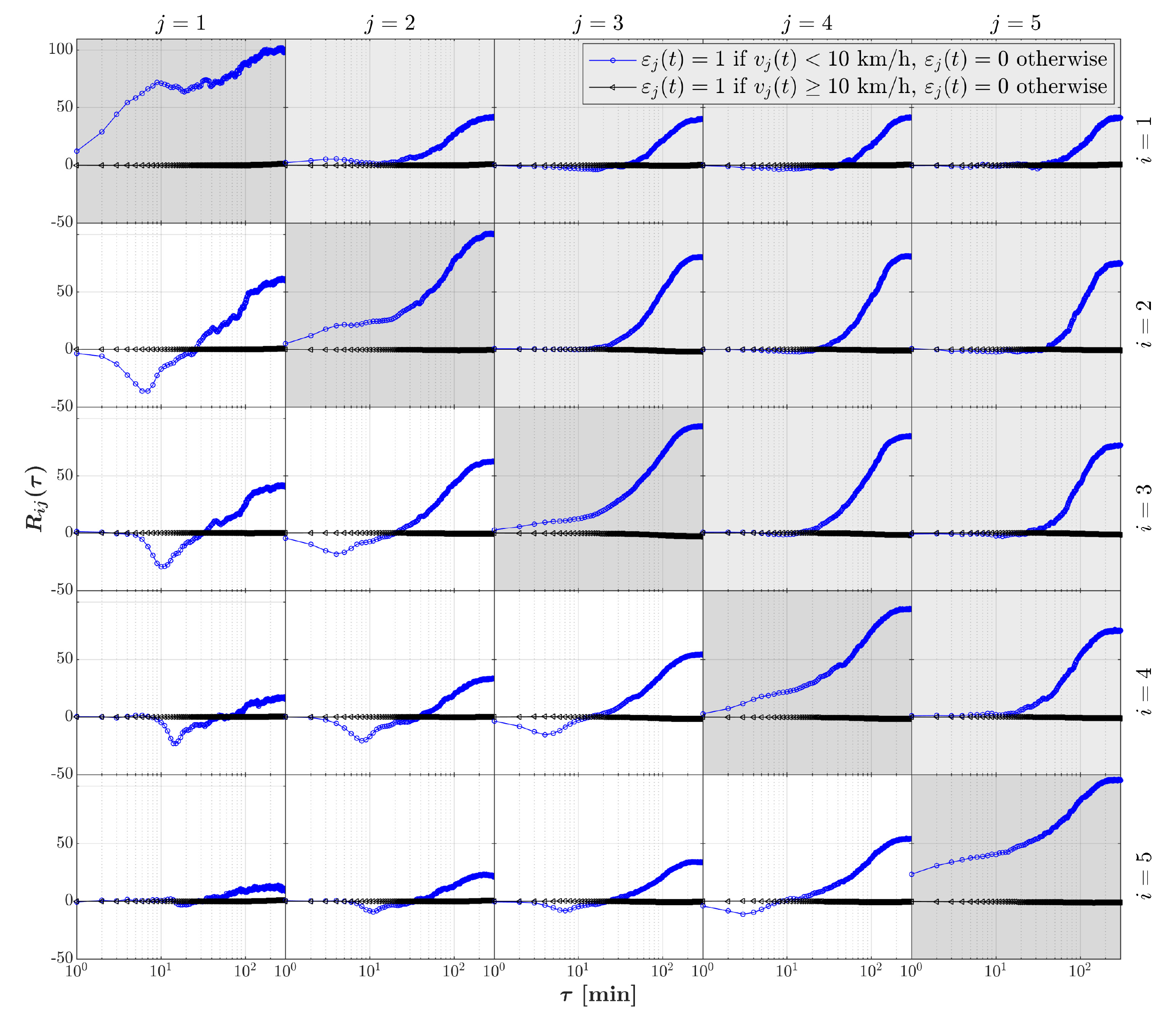}
\caption{The velocity response $R_{ij}(\tau)$ versus time lag $\tau$ in a single logarithmic scale. The upper (lower) triangle subplots with light grey (white) background correspond to the forward (backward) propagation of heavy congestion of section $j$, where $j=1,~\cdots~5$, respectively, for each columns of subplots.}
\label{fig3}
\end{center}
\vspace*{-0.5cm}
\end{figure}

Unlike other complex systems, a traffic network does not only involve time and space scales, but might also be differentiated by directions of traffic flow. Thus, it makes a difference if the directions of flow propagation and of the sequence of considered sections are parallel or antiparallel. For a heavily congested section on a motorway, the congestion might propagate forward along or backward against the direction of traffic flow, regardless of the extent of effects. Both the forward and the backward sections might respond to the section with heavy congestion. This inference prompts us to observe five sequential sections on a motorway. Figure~\ref{fig3} shows the backward (forward) propagation of heavy congestion on section 1 (section 5) and both the backward and forward propagation of heavy congestion on sections 2, 3 and 4. For each of congested sections, the velocity responses from its neighbouring sections are visible.

The response to heavy congestion is strong for each section. In particular, during a short time period of some dozens of minutes, the response behavior varies with the position of the responding section $i$ relative to the congested section $j$. As for the heavy congestion with backward propagation from section 1, the transient response during a short time period shows a velocity dip down first and then a recovery. The recovered velocity continues to increase with time, enhancing the velocity response from dozens of minutes to 300 minutes. The amplitude of the transient response, however, shrinks the farther the responding section $i$ is from the congested section $j$. On the other hand, for heavy congestion with the forward propagation from section 5, the dip down behavior of the transient response disappears. Instead, the transient response is close to zero during several dozens of minutes. These two kinds of behavior for transient responses with and without dipping down are significantly different and can be observed separately in the backward and forward propagation of heavy congestion on sections 2, 3 and 4 in figure~\ref{fig3}.

With regard to Kerner's three-phase traffic theory~\cite{kerner2021}, the wide moving jam is approximately related to the vehicle velocity lower than 10 km/h, which is exactly the case of $\varepsilon_j(t)=1$ in our study. Therefore, for a motionless bottleneck with a wide moving jam, e.g., section 1, the dip down behavior with the increase of amplitude in velocity responses of sections from 5 to 2 in figure~\ref{fig3} reflects a transition of traffic phases from free flow to the wide moving jam along with the vehicle deceleration. For section 5 as the bottleneck with a wide moving jam, the absence of dip down behavior in transient responses of sections from 4 to 1 very likely manifests the escape of vehicles via acceleration from the wide moving jam to free flow or synchronized flow. Consequently, the distinct response behavior taking into account the backward and forward propagation of heavy congestion reveals the different transitions of traffic phases.

\begin{figure}[tb]
\begin{center}
\includegraphics[width=1\textwidth]{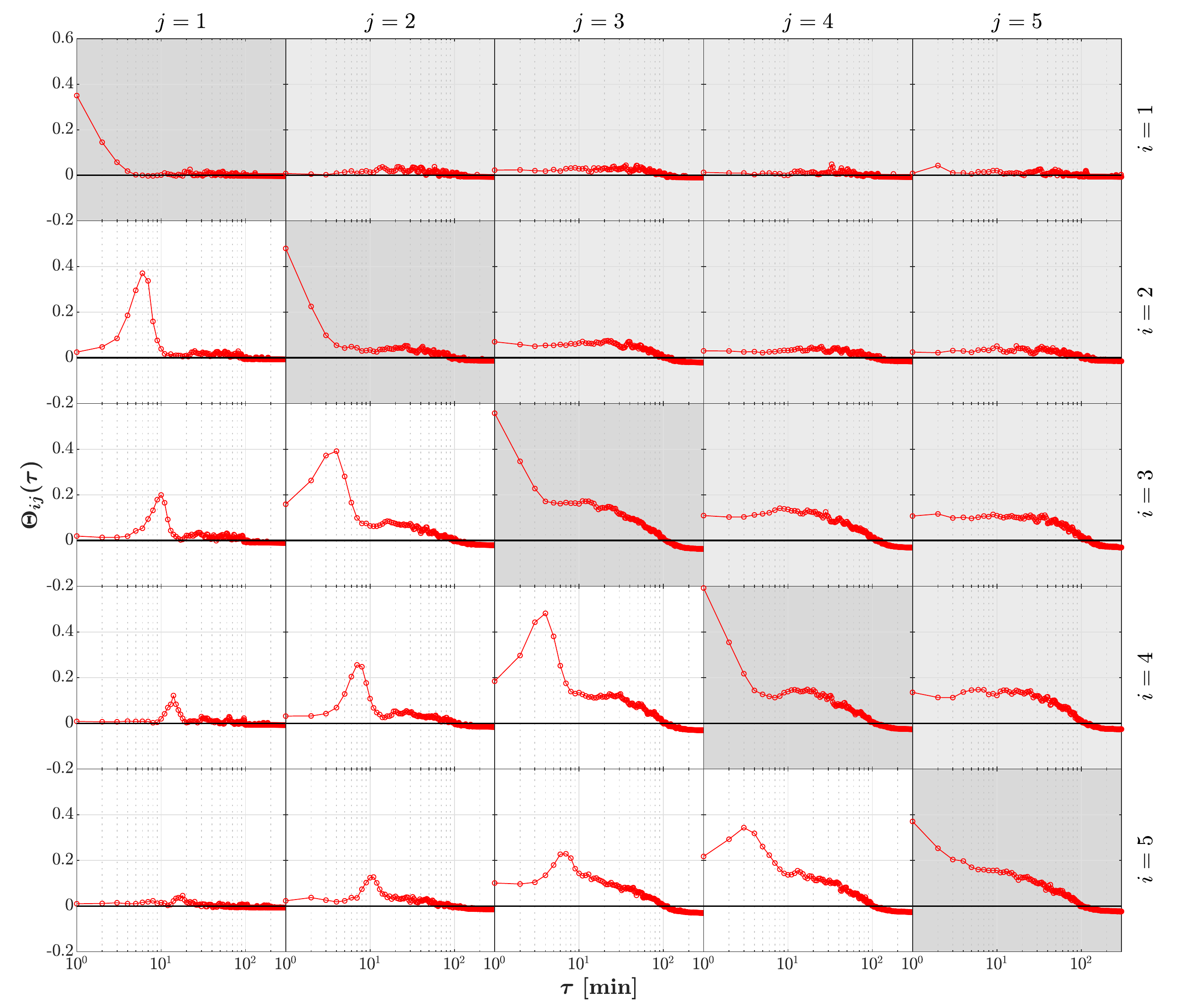}
\caption{The congestion correlator $\Theta_{ij}(\tau)$ versus time lag $\tau$ in a single logarithmic scale. The upper (lower) triangle subplots with light grey (white) background correspond to the forward (backward) propagation of heavy congestion of section $j$, where $j=1,~\cdots~5$, respectively, for each columns of subplots.}
\label{fig4}
\end{center}
\vspace*{-0.5cm}
\end{figure}

\subsection{Separating response phases}
\label{sec33}

As seen, the response functions $R_{ij}(\tau)$ in figure~\ref{fig3} and the congestion correlators $\Theta_{ij}(\tau)$ in figure~\ref{fig4} are clearly related. For heavy congestion with backward propagation, accompanied with a valley in the transient response, a corresponding peak is visible in the congestion correlator. As for heavy congestion with forward propagation, if the transient response tends to zero, the congestion correlator basically keeps constant with time lag $\tau$. In both cases, again, the relative position of congested sections produces an effect on the amplitude of the congestion correlator during a short time period. The farther separated the two congested sections are, the weaker the amplitude of their correlation is. This might be because the ramps and crossroads along the route between two farther separated sections relieve the effects of heavy congestion on a section that is far away from the congested section. In contrast, the short distance between two sections limits the possibility to effectively reduce the effect of heavy congestion on the section that is close to the congested section, giving rise to a strong correlation of heavy congestion between them. 
 
 \begin{figure}[tb]
\begin{center}
\includegraphics[width=0.48\textwidth]{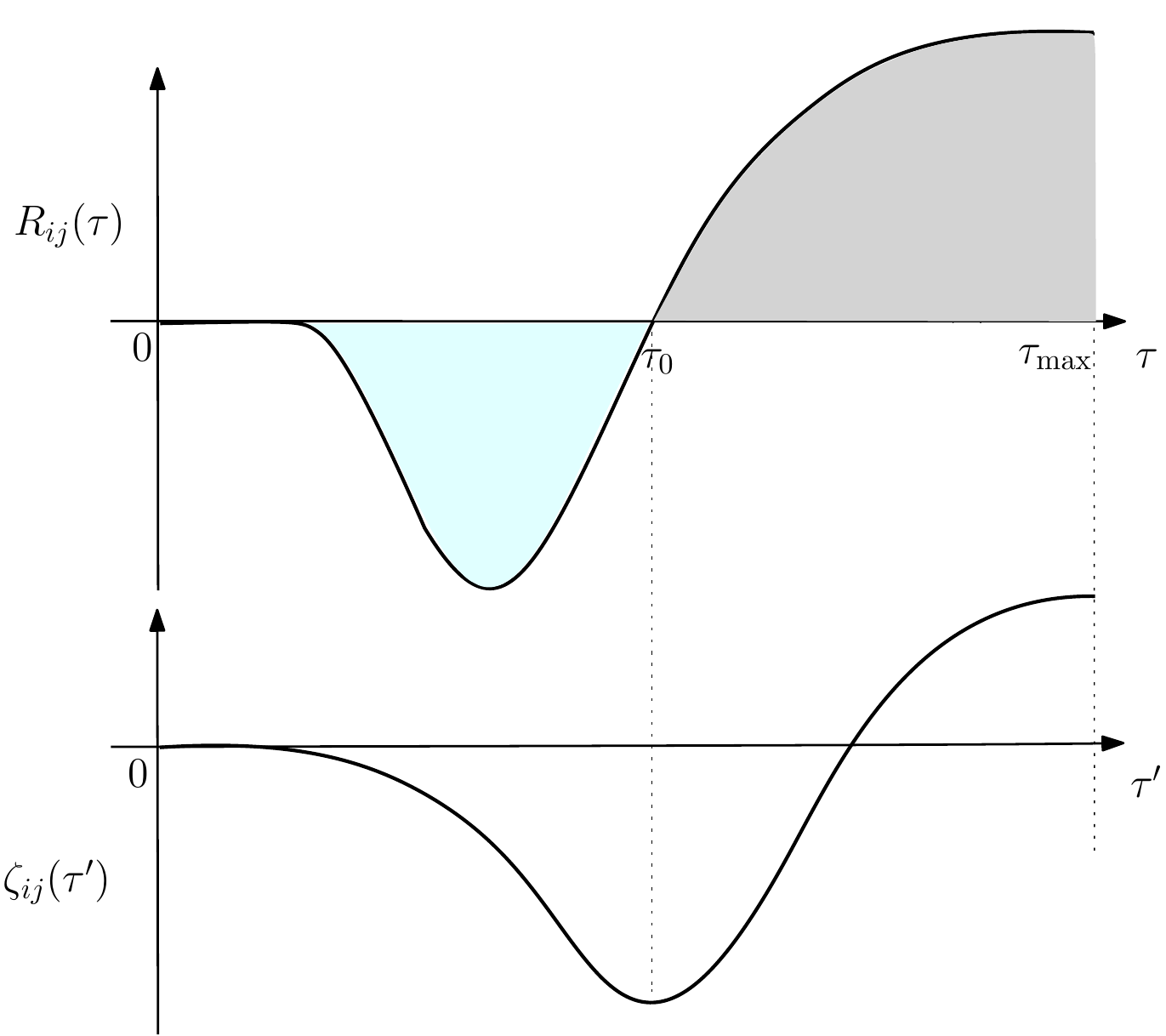}
\includegraphics[width=0.48\textwidth]{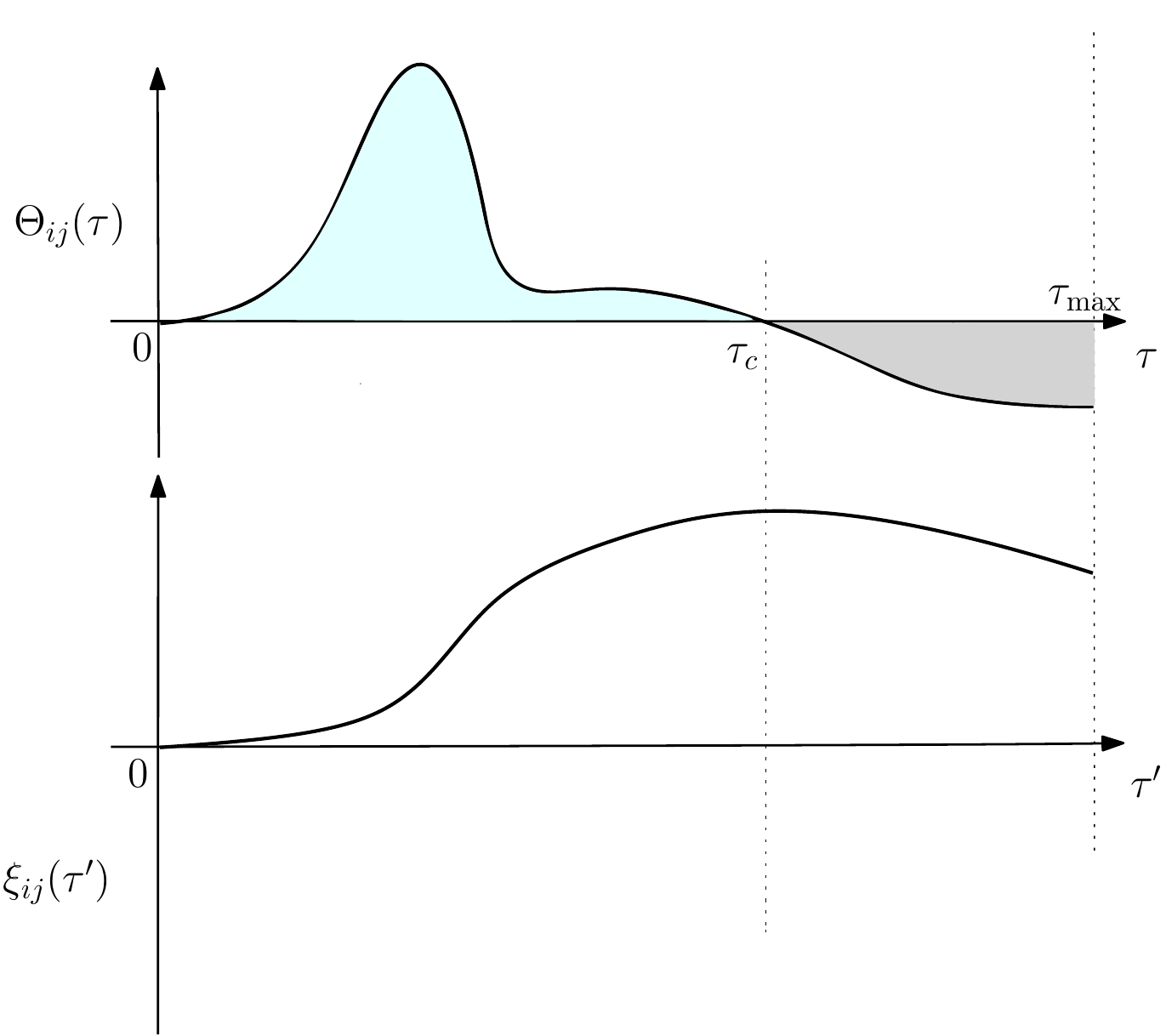}
\caption{Left: a sketch of $R_{ij}(\tau)$ versus $\tau$ and accordingly $\zeta_{ij}(\tau')$ versus $\tau'$; right: a sketch of $\Theta_{ij}(\tau)$ versus $\tau$ and accordingly $\xi_{ij}(\tau')$ versus $\tau'$. The coloured areas are the integral of a quantity, i.e., $R_{ij}(\tau)$ or $\Theta_{ij}(\tau)$, over $\tau$. The light blue area is the integral, i.e., $\zeta_{ij}(\tau_0)$ or $\xi_{ij}(\tau_c)$, of a quantity, i.e., $R_{ij}(\tau)$ or $\Theta_{ij}(\tau)$, over $\tau$ from 0 to a critic time, i.e., $\tau_0$ or $\tau_c$, where the critical time $\tau_0$ and $\tau_c$ in the curves of $R_{ij}(\tau)$ and $\Theta_{ij}(\tau)$ correspond to the minimum and maximum of the curves of $\zeta_{ij}(\tau')$ and $\xi_{ij}(\tau')$, respectively.
}
\label{fig5}
\end{center}
\vspace*{-0.5cm}
\end{figure}

\begin{figure}[tb]
\begin{center}
\includegraphics[width=1\textwidth]{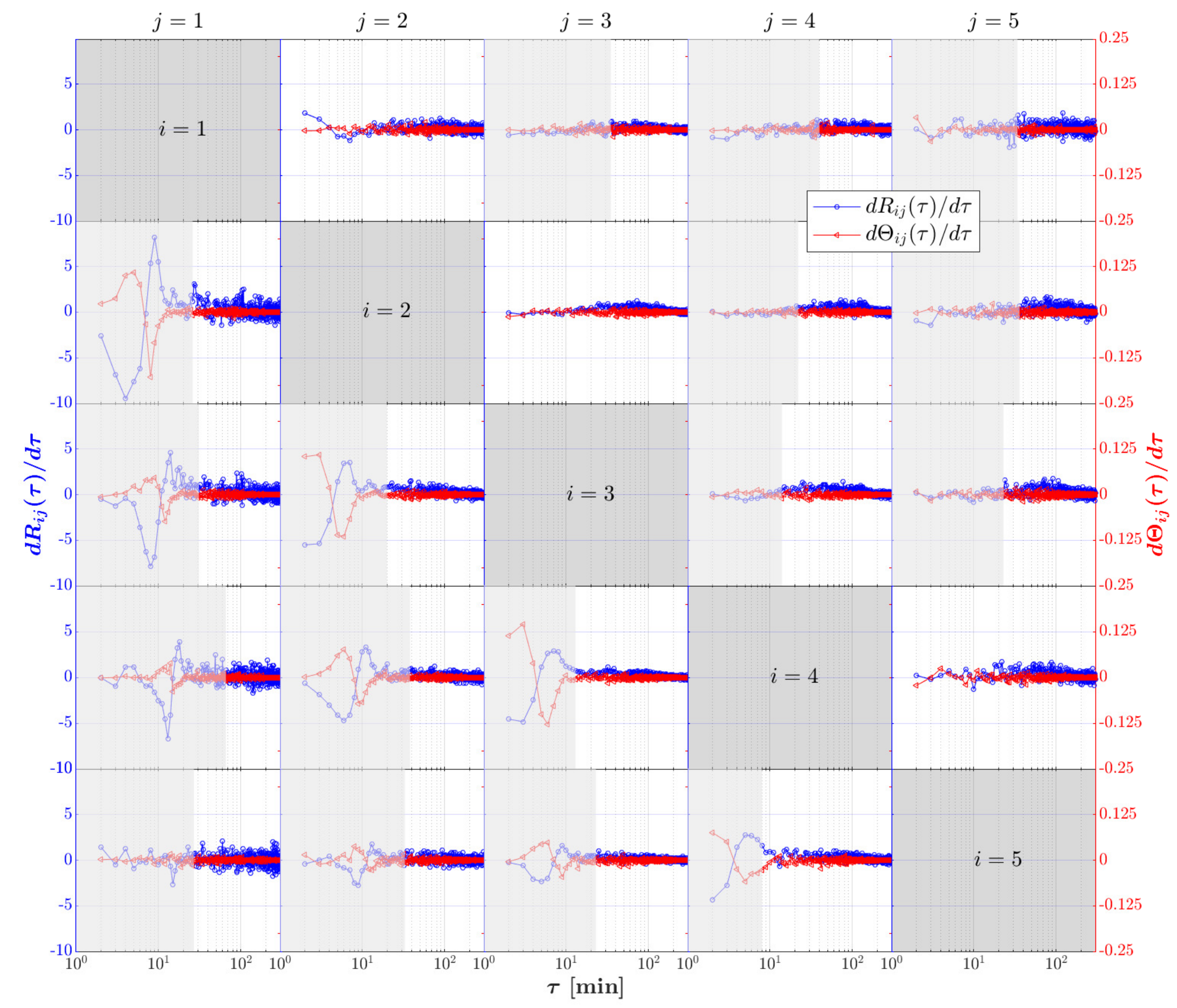}
\caption{The dependences of the derivatives $dR_{ij}(\tau)/d\tau$ (corresponding to the left vertical axis) and  $d\Theta_{ij}(\tau)/d\tau$ (corresponding to the right vertical axis) on time lag $\tau$ in a single logarithmic scale. The region with light grey in each subplot covers the duration of response phase 1. The upper (lower) triangle subplots correspond to the forward (backward) propagation of heavy congestion of section $j$, where $j=1,~\cdots,~5$, respectively, for each row of subplots.}
\label{fig6}
\end{center}
\vspace*{-0.5cm}
\end{figure}

\begin{figure}[tb]
\begin{center}
\includegraphics[width=1\textwidth]{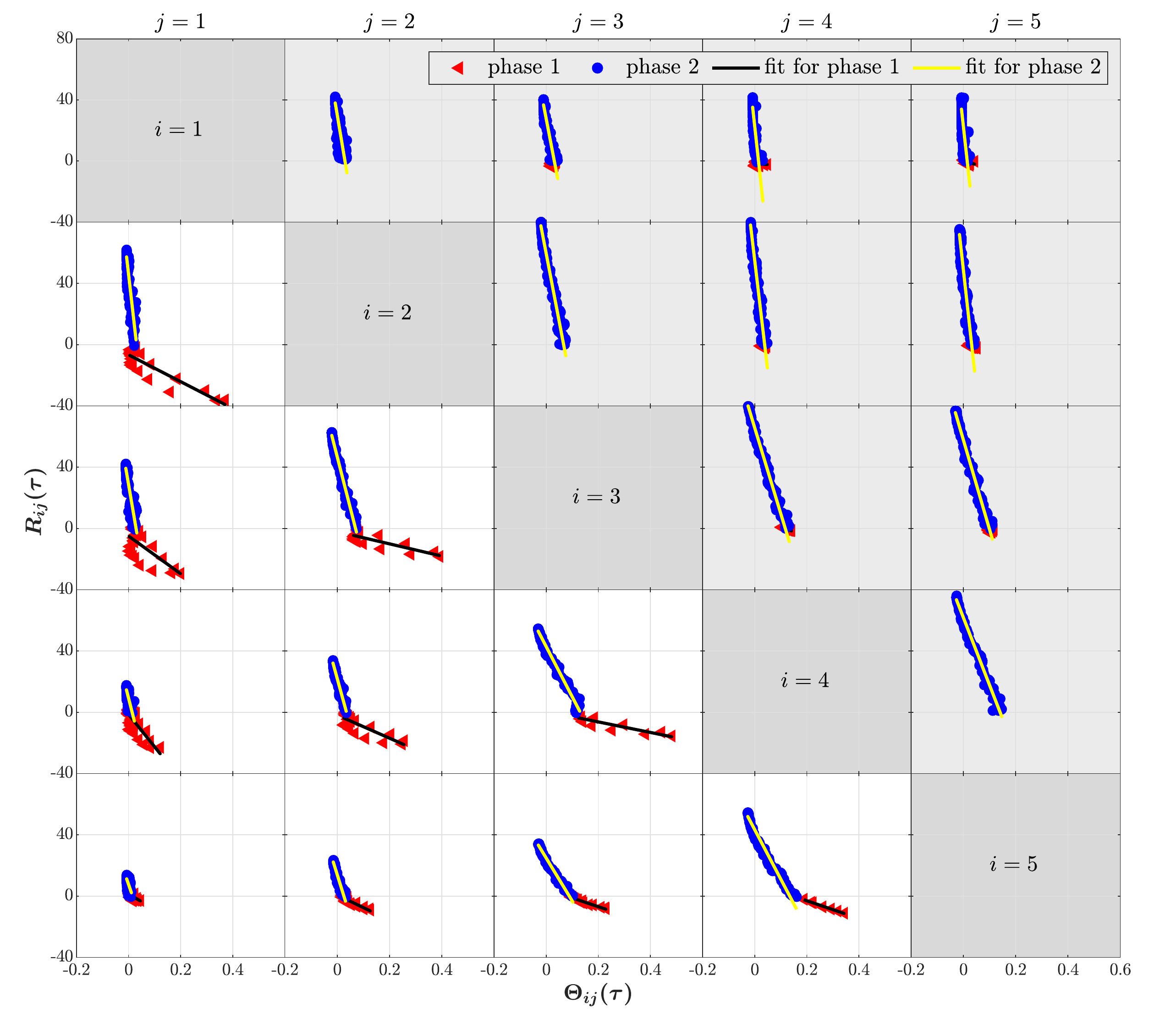}
\caption{Scatter plots of velocity response $R_{ij}(\tau)$ versus congestion correlator $\Theta_{ij}(\tau)$. The upper (lower) triangle subplots with light grey (white) background correspond to the forward (backward) propagation of heavy congestion of section $j$, where $j=1,~\cdots,~5$, respectively, for each row of subplots. The points of each response phase in each subplot are fitted by a first-degree linear polynomial curve.}
\label{fig7}
\end{center}
\vspace*{-0.5cm}
\end{figure}

Figures~\ref{fig3} and \ref{fig4} feature a connection between congestion correlators and velocity responses. As mentioned above, the velocity responses comprise a phase of transient responses and a phase of long-term responses. To quantify the relation between the congestion correlator and the velocity response during the first dozens of minutes, a separation of the two response phases is called for. The behavior of dipping down and recovering in the transient response is analogous to traffic resilience~\cite{Bruneau2003,Tang2018,Zhang2019}, which might be by a function in terms of argument $\tau'$ defined as
\begin{equation}
\zeta_{ij}(\tau')=\int\limits_0^{\tau'} R_{ij}(\tau)d\tau=\sum\limits_{\tau=0}^{\tau'}R_{ij}(\tau)\Delta \tau
\label{eq34}
\end{equation}
with $\Delta\tau=1$ min. Here $\tau'$ is the time lag to which the response is integrated, and $\tau_0$ is the critical time at which the velocity recovers to the previous value before time lag, i.e., at $\tau=0$. As schematically depicted in figure~\ref{fig5}, when the response $R_{ij}(\tau)$ changes from the previous continuous negative values to continuous positive values, $\zeta_{ij}(\tau')$ reaches the minimum, because after the critical time $\tau_0$, only positive response values add to the integral, leading to an increase of $\zeta_{ij}(\tau')$ with $\tau'$. Therefore, the minimal value of $\zeta_{ij}(\tau')$ implies the critical time $\tau_0$ such that
\begin{equation}
\tau_0=\mathop{\mathrm{argmin}}_{\tau'\in[1,~\tau_\mathrm{max}]}\zeta_{ij}(\tau') \ ,
\label{eq35}
\end{equation}
where $\tau_\mathrm{max}=300$ min in this study. We use the obtained critical time $\tau_0$ to separate the two response phases.

We further explore the changes in the velocity responses by working out the difference of responses between two neighbouring time lags , 
\begin{equation}
\frac{dR_{ij}(\tau)}{d\tau}=\frac{R_{ij}(\tau+\Delta \tau)-R_{ij}(\tau)}{\Delta \tau} \ ,
\label{eq36}
\end{equation}
analogously for the changes in the congestion correlators, 
\begin{equation}
\frac{d\Theta_{ij}(\tau)}{d\tau}=\frac{\Theta_{ij}(\tau+\Delta \tau)-\Theta_{ij}(\tau)}{\Delta \tau} \ ,
\label{eq37} 
\end{equation}
with $\Delta \tau=1$ min. The two kinds of changes, both shown in the figure~\ref{fig6}, are partitioned into two response phases by the critical time $\tau_0$. During the first response phase, the velocity responses and congestion correlators fluctuate dramatically and reversely, causing this phase unstable. In contrast, during the second response phase, the two quantities fluctuate mildly and their changes tend to $dR_{ij}(\tau)/d\tau\approx0$ and $d\Theta_{ij}(\tau)/d\tau\approx0$, implying a nearly stable phase. The difference in the stability makes the first phase corresponding to the transient response more interesting and valuable to explore. As shown in figure~\ref{fig7}, during each response phase $k$ (RP $k$, $k=1$ or 2), the velocity response is inversely proportional to the congestion correlator, 
\begin{equation}
R_{ij}(\tau)|_{\mathrm{RP~}k}\sim -\Theta_{ij}(\tau)|_{\mathrm{RP~}k} \ ,
\label{eq38}
\end{equation}
in particular for the second phase which is more stable. This relation suggests that the congestion correlator between neighbouring sections may be viewed as a major reason for the velocity responses, irrespective of other causes.

\subsection{Duration of heavy congestion}
\begin{figure}[tb]
\begin{center}
\includegraphics[width=1\textwidth]{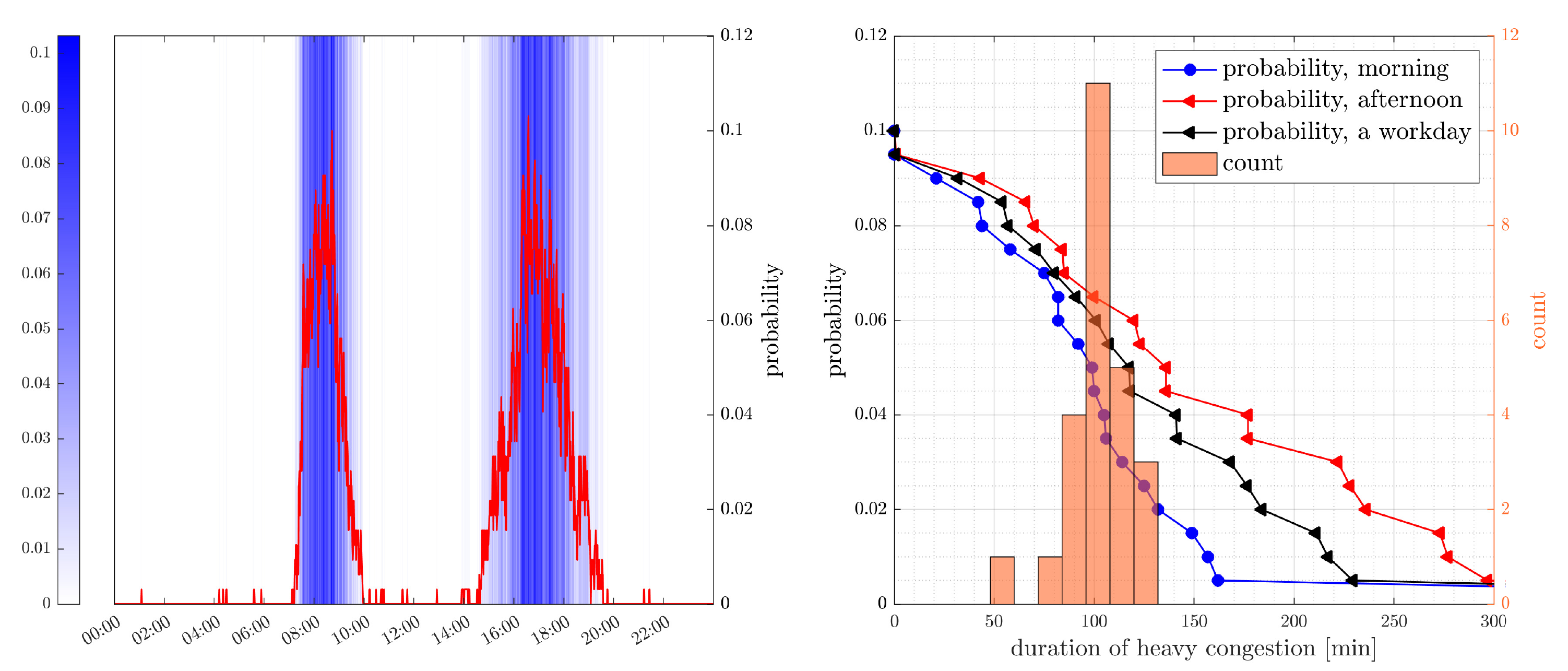}
\caption{Left: time evolution of empirically occurring probability of heavy congestion during a workday, where the blue color highlights the value of probability; right: the empirical occurring probabilities in the morning, in the afternoon and in a whole workday versus the empirical duration of heavy congestion, as well as the distribution of theoretical duration of heavy congestion worked out by equations~\eqref{eq39} and \eqref{eq40}.}
\label{fig8}
\end{center}
\vspace*{-0.5cm}
\end{figure}

We notice that for each case in figure~\ref{fig4}, the correlation of heavy congestion vanishes at a similar value around 100 min, regardless of self- or cross-correlators. This critical time $\tau_c$ for $\Theta_{ij}(\tau_c)=0$ very likely suggests the duration of heavy congestion that contributes to the congestion correlators. To determine $\tau_c$, we employ a method similar to equation~\eqref{eq34} by considering a resilience function of time lag $\tau'$
\begin{equation}
\xi_{ij}(\tau')=\int\limits_0^{\tau'} \Theta_{ij}(\tau)d\tau=\sum\limits_{\tau=0}^{\tau'}\Theta_{ij}(\tau)\Delta \tau \ ,
\label{eq39}
\end{equation}
where $\Delta \tau=1$ min and $\tau_\mathrm{max}=300$ min. As depicted in figure~\ref{fig5}, when $\xi_{ij}(\tau')$ reaches its maximum, the congestion correlator crosses zero, $\Theta_{ij}(\tau_c)=0$, since the subsequent negative values of $\Theta_{ij}(\tau)$ will reduce the value of $\xi_{ij}(\tau')$. Therefore, we can obtain the critical time $\tau_c$ by
\begin{equation}
\tau_c=\mathop{\mathrm{argmax}}_{\tau'\in[1,~\tau_\mathrm{max}]}\xi_{ij}(\tau') \ ,
\label{eq40}
\end{equation}
listed in table~\ref{tab1} and visualized in the histogram in figure~\ref{fig8} (right). As observed in figure~\ref{fig4}, most of the critical times $\tau_c$ are around 100 min. To quantify this time scale, we average the time series of indicators $\varepsilon_j(t)$ over five sections and 64 considered workdays. The resulting time series, shown in figure~\ref{fig8} (left), indicate the empirically occurring probability of heavy congestion at each time $t$. Above each given threshold of occurring probabilities, we display the largest empirical duration of heavy congestion in figure~\ref{fig8} (right). The result reveals that 
heavy congestion is more likely to occur with a short duration. According to reference~\cite{Krause2017}, short-lasting traffic jams, e.g. jams with duration smaller or equal to 10 min (contributing $19\%$ to the total sum of jam hour), yield a large risk for rear-end collisions, while the long durations are of minor importance. Here we find the average occurring probability of heavy congestion at a time $t$ with short duration smaller or equal to 10 min is $19.5\%$, implying a high risk for rear-end collisions. The congestion correlation, however, comes from the long-lasting heavy congestion with duration around 100 min. The overlapping between the bar around 100 min in the histogram and the three curves in figure~\ref{fig8} (right) discloses that the average occurring probability is $3.5\%\sim 5\%$ for the morning, $6.5\%$ for the afternoon and $5.5\%\sim 6\%$ for a whole workday.

\begin{table}[tb]
\caption{The critical time $\tau_c$ corresponding to figure~\ref{fig4} }
\vspace*{-0.3cm}
\begin{center}
\setlength{\tabcolsep}{21pt}
\begin{tabular*}{0.8\textwidth}{c|ccccc}
\hlineB{2}
\diagbox{$i$}{$j$}& 1 & 2 & 3 & 4 & 5 \\
\hline
1&\cellcolor{darkgray}{79}      &\cellcolor{lightgray}{109}   &\cellcolor{lightgray}{105}    &\cellcolor{lightgray}{99}    &\cellcolor{lightgray}{118}\\
2&92    &\cellcolor{darkgray}{91}    &\cellcolor{lightgray}{105}    &\cellcolor{lightgray}{122}    &\cellcolor{lightgray}{120}\\
3&95    &105    &\cellcolor{darkgray}{107}    &\cellcolor{lightgray}{108}   &\cellcolor{lightgray}{124}\\
4&98    &100    &108 &\cellcolor{darkgray}{106}      &\cellcolor{lightgray}{113}\\
5&54    &86    &100  &103     & \cellcolor{darkgray}{100}\\
\hlineB{2}
\end{tabular*}
\end{center}
\label{tab1}
\end{table}%

\section{Dynamic description of the transient response}
\label{sec4}

The propagation of heavy congestion in opposite directions gives rise to different dynamic behavior in velocity responses. Here we focus on the transient response in the first dozens of minutes after a heavy congestion sets in. In section~\ref{sec41}, we construct a susceptible-decelerated-withdrawing (SDW) model to describe the transient response. In section~\ref{sec42}, we reveal the local dynamic mechanism of transient responses through the parameters of propagation rates and recovery rates from this model.

\subsection{Susceptible-decelerated-withdrawing model}
\label{sec41}

\begin{figure}[htbp]
\begin{center}
\includegraphics[width=1\textwidth]{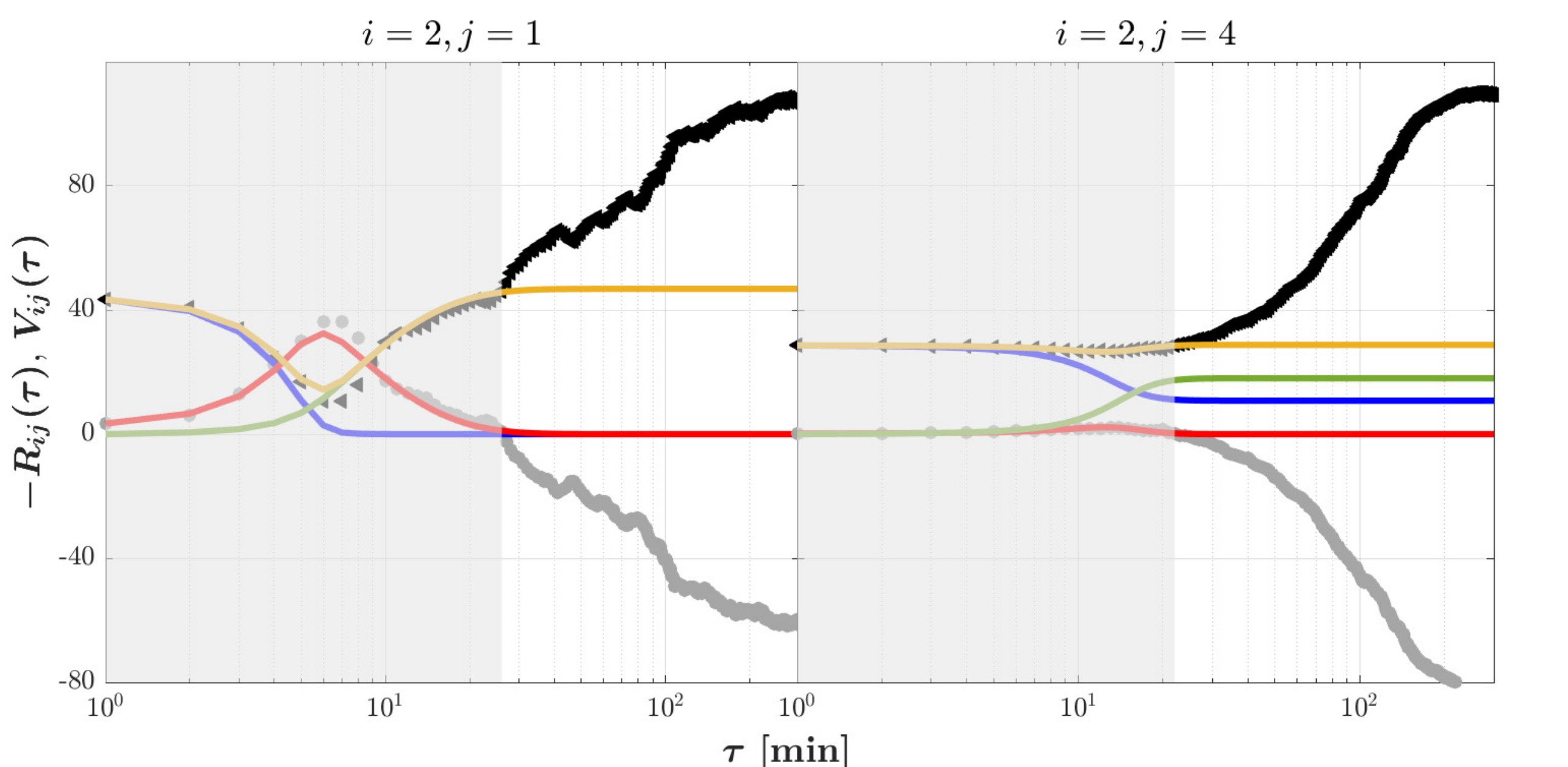}
\includegraphics[width=1\textwidth]{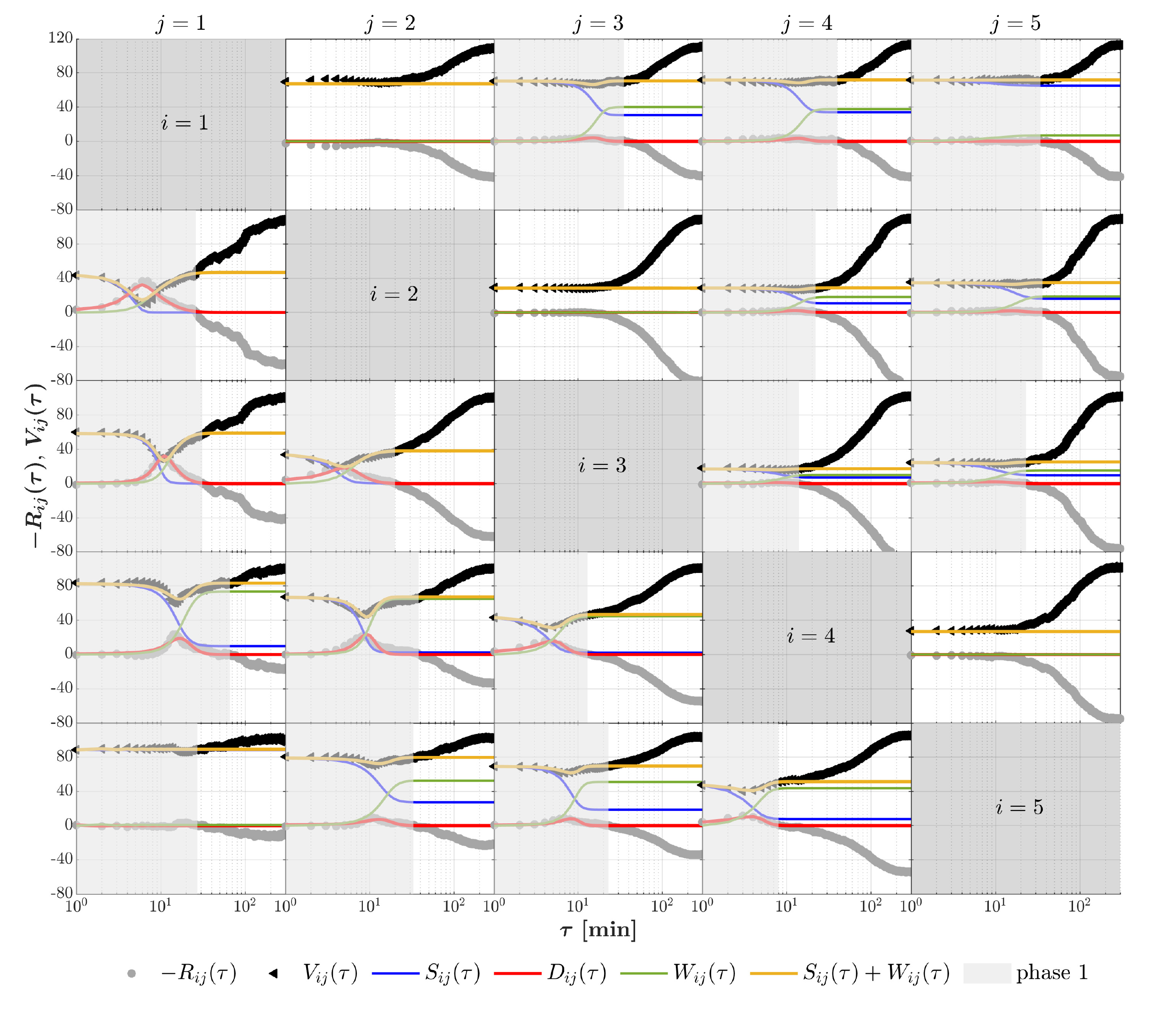}
\caption{The empirical $-R_{ij}(\tau)$ and  $V_i(\tau)|_j$ versus time lag $\tau$, fitted by the SDW model. The region with light grey in each subplot covers the duration of phase 1. In the $5\times5$ subplot matrix, the upper (lower) triangle subplots correspond to the forward (backward) propagation of heavy congestion of section $j$, where $j=1,~\cdots,~5$, respectively, for each row of subplots. The first two subplots are examples  zoomed in for details.}
\label{fig9}
\end{center}
\end{figure}

The unstable characteristics in the transient response make the dynamic mechanism of responses interesting for further study. We reformulate the velocity change of section $i$ due to the heavy congestion of section $j$, i.e., the response function~\eqref{eq33}, by
\begin{equation}
R_{ij}(\tau)=V_{ij}(\tau)-V_{ij}(0) \ ,
\label{eq41}
\end{equation}
where $V_{ij}(\tau)$ is the remaining velocity of section $i$, on average, subject to the effect of heavy congestion on section $j$, and $V_{ij}(0)$ is the average initial velocity of section $i$ at $\tau=0$ without the effect of heavy congestion, i.e., 
\begin{equation}
V_{ij}(\tau) =\frac{\sum\limits_{t=1}^{T-\tau}v_i(t+\tau)\varepsilon_j(t)}{\sum\limits_{t=1}^{T-\tau}\varepsilon_j(t)}  \quad\mathrm{and}\quad V_{ij}(0)=\frac{\sum\limits_{t=1}^{T-\tau}v_i(t)\varepsilon_j(t)}{\sum\limits_{t=1}^{T-\tau}\varepsilon_j(t)} \ .
\label{eq42}
\end{equation}
Obviously, $V_{ij}(\tau)$ changes with time lag $\tau$. In contrast, $V_{ij}(0)$ is independent of time lag $\tau$ and stays constant. At any time lag during the phase of transient responses, it amounts to the sum of the average decelerated velocity $-R_{ij}(\tau)$ and the average remaining velocity $V_{ij}(\tau)$,
\begin{equation}
V_{ij}(0)=-R_{ij}(\tau)+V_{ij}(\tau)\ .
\label{eq44}
\end{equation}
While $-R_{ij}(\tau)$ increases to a certain extent with $\tau$, $V_{ij}(\tau)$ decreases accordingly with $\tau$. Their $\tau$-dependent curves are symmetrical in each subplot of figure~\ref{fig9}. As described in reference~\cite{kerner2021}, the transition from free flow to synchronized flow to free flow (F$\rightarrow$S$\rightarrow$F transition) at a bottleneck is due to the competition of driver speed adaptation and over-acceleration. Analogously, we assume the velocity during the phase of transient responses results from the competition of two causes. One is the driver speed adaptation on section $i$ after the influence of heavy congestion on section $j$. This speed adaptation leads to a susceptible velocity $S_{ij}(\tau)$ which would decay via vehicle deceleration from $V_{ij}(0) $ to a non-negative value with $\tau$. The other one is the driver over-acceleration on section $i$ for recovering from the influence of heavy congestion on section $j$. This over-acceleration gives rise to a withdrawing velocity $W_{ij}(\tau)$ which would raise from 0 to a value not higher than $V_{ij}(0)$ with $\tau$. Therefore, we can further decompose the average remaining velocity into a susceptible velocity $S_{ij}(\tau)$ and a withdrawing velocity $W_{ij}(\tau)$. 
 \begin{equation}
V_{ij}(\tau) =S_{ij}(\tau)+W_{ij}(\tau) \ .
 \label{eq45}
\end{equation}
We find it convenient to introduce the decelerated velocity
\begin{equation}
D_{ij}(\tau)=-R_{ij}(\tau) \ ,
 \label{eq455}
\end{equation}
implying that equation~\eqref{eq44} takes the form
\begin{equation}
 V_{ij}(0)=D_{ij}(\tau)+S_{ij}(\tau)+W_{ij}(\tau) \ .
 \label{eq46}
\end{equation}
Therefore, the average velocity without any time lag $V_{ij}(0)$ is the balance of the susceptible velocity $S_{ij}(\tau)$, the decelerated velocity $D_{ij}(\tau)$ and the withdrawing velocity $W_{ij}(\tau)$. Quite remarkably, equation~\eqref{eq46} is formally a conservation law: there is no velocity leaving or entering this ``velocity system".

The susceptible-infectious-recovered (SIR) model~\cite{Kermack1927} is one of the simplest models in epidemiology, 
\begin{eqnarray}
&&\frac{dS(t)}{dt} = -\frac{\beta I(t)S(t)}{N} \label{eq461} \ , \\   [6pt]
&&\frac{dI(t)}{dt} = \frac{\beta I(t)S(t)}{N} -\gamma I(t)  \label{eq462} \ , \\   [6pt]
&&\frac{dR(t)}{dt} = \gamma I(t)   \label{eq463} \ ,
\end{eqnarray}
where the total population $N$, including the number of susceptible individuals $S(t)$,  the number of infectious individuals $I(t)$ and the number of recovered individuals $R(t)$, 
\begin{equation}
N=S(t)+I(t)+R(t) \ ,
\label{eq47}
\end{equation}
is constant without considering the dynamics of birth and death. Obviously, equations~\eqref{eq46} and \eqref{eq47} formally correspond to each other.

The SIR model has been applied to describing the propagation of congestion in urban traffic or in airspace~\cite{Zeng2018,Fan2020,Dai2016,Saberi2020}. Inspired by this model, we construct a susceptible-decelerated-withdrawing (SDW) model to describe the transient response, 
 \begin{eqnarray}
&& \frac{dS_{ij}(\tau)}{d\tau} =\frac{S_{ij}(\tau+\Delta \tau)-S_{ij}(\tau)}{\Delta \tau} = -\frac{\beta_{ij} D_{ij}(\tau)S_{ij}(\tau)}{V_{ij}(0)}  \  , \label{eq471} \\ [6pt] 
&& \frac{dD_{ij}(\tau)}{d\tau} =\frac{D_{ij}(\tau+\Delta \tau)-D_{ij}(\tau) }{\Delta \tau}= \frac{\beta_{ij} D_{ij}(\tau)S_{ij}(\tau)}{V_{ij}(0)}-\gamma_{ij} D_{ij}(\tau)  \ , \label{eq472} \\ [6pt]  
&& \frac{dW_{ij}(\tau)}{d\tau} =\frac{W_{ij}(\tau+\Delta \tau)-W_{ij}(\tau) }{\Delta \tau}=\gamma_{ij} D_{ij}(\tau)   \ .  \label{eq473}  
 \end{eqnarray}
Here $\Delta \tau=1$ min. Apparently, the above model does not contain any visible variable about space, e.g., distances, but the space information is incorporated into the two parameters $\beta_{ij}$ and $\gamma_{ij}$ and influences their magnitude. To study the dynamic mechanism of transient responses that are influenced by local space information, we consider the decelerated velocity in the model as a spatiotemporal function in terms of $\tau$, $\beta_{ij}$ and $\gamma_{ij}$, represented by $D_{ij}(\tau | \beta_{ij},\gamma_{ij})$. The residual sum of squares (RSS)~\cite{Archdeacon1994} which measures the discrepancy between the estimated $D_{ij}(\tau | \beta_{ij},\gamma_{ij})$ and the empirical decelerated velocity, i.e., the empirical negative response $-R_{ij}(\tau)$, reads
 \begin{equation}
\Xi(\beta_{ij},\gamma_{ij})=\sum\limits_{\tau=1}^{\tau_\mathrm{max}} (-R_{ij}(\tau)-D_{ij}(\tau | \beta_{ij},\gamma_{ij}))^2 \ .
\label{eq48}
 \end{equation}
In order to find the optimal parameters $\beta_{\ij,\mathrm{fit}}$ and $\gamma_{ij,\mathrm{fit}}$, we apply the least square fitting to $D_{ij}(\tau | \beta_{ij},\gamma_{ij})$ (simulated by equations~\eqref{eq471}, \eqref{eq472} and \eqref{eq473}) and the empirical $-R_{ij}(\tau)$ by minimizing the RSS $\Xi(\beta_{ij},\gamma_{ij})$, 
 \begin{equation}
\{\beta_{ij,\mathrm{fit}},\gamma_{ij,\mathrm{fit}}\}=\mathop{\mathrm{argmin}}_{\beta_{ij}\in[0,~2],\gamma_{ij}\in[0,~2]} \ \Xi(\beta_{ij},\gamma_{ij}) \  
\label{eq49}
\end{equation}
within the given ranges $\beta_{ij}\in[0,~2]$ and $\gamma_{ij}\in[0,~2]$.

The simulation of the model~\eqref{eq471}, \eqref{eq472} and \eqref{eq473} requires initial values $S_{ij}(0)$, $D_{ij}(0)$ and $W_{ij}(0)$ as input. A negative value of responses during the first $\tau_\mathrm{max}$ suggests a negative effect of heavy congestion of section $j$ propagating to section $i$, leading to the deceleration of vehicles. To include this effect which drives the dynamic evolution of velocity, we set the initial value $D_{ij}(0)$ approximating to the additive inverse of the first negative value of responses, i.e., approximating to the first $-R_{ij}(\tau)>0$, in the time range $1\leq \tau \leq\tau_\mathrm{max}$ with $\tau_\mathrm{max}=300$ min. Without any negative value of responses detected during the time range, the dynamic evolution of velocity is absent and $D_{ij}(0)=0$ in this case. We also set $W_{ij}(0)=0$ since the non-zero withdrawing velocity lags behind the initial deceleration, and $S_{ij}(0)=V_{ij}(0)-D_{ij}(0)-W_{ij}(0)$ according to equation~\eqref{eq46}.

\subsection{Propagation and recovery rates}
\label{sec42}

Figure~\ref{fig5}  shows the fitted $S_{ij}(\tau)$ and the fitted $W_{ij}(\tau)$, allowing for a comparison of the differences between the empirical $-R_{ij}(\tau)$ and the fitted $D_{ij}(\tau)$ as well as between the empirical $V_{ij}(\tau) $ and the fitted $S_{ij}(\tau)+W_{ij}(\tau)$. The fitted and the empirical values match well during the phase of transient responses, i.e., during the unstable phase of responses. In the phase of long-term response they differ, as our model is not supposed to capture this phase. 

We take a closer look at the resulting parameters. Tables~\ref{tab2}, \ref{tab3} and \ref{tab4} list the optimal parameters $\beta_{\ij,\mathrm{fit}}$, $\gamma_{ij,\mathrm{fit}}$ and their ratio $\beta_{ij,\mathrm{fit}}/\gamma_{ij,\mathrm{fit}}$, respectively. The grey (white) cells in the three tables correspond to the forward (backward) propagation of heavy congestion of section $j$. The zero values in the tables are due to the absence of dynamic evolution in transient responses with $D_{ij}(0)=0$. Table~\ref{tab2} yields that the propagation rates $\beta_{ij,\mathrm{fit}}$ are very close in the two cases of propagation, implying that the effects of heavy congestion in a section on the velocity in its neighbouring sections are similar whether it propagates forward or backward. In contrast, the difference of the recovery rates $\gamma_{ij,\mathrm{fit}}$ for the two cases of propagation is evident in table~\ref{tab3}. Roughly, the rates for velocity recovery in the case of forward propagation are larger than those in the case of backward propagation. The absence of congested traffic flow in front of a section facilitates the vehicle acceleration and make the velocity easily recover to the value before the effect from forward propagation of heavy congestion. In other words, a heavily congested traffic flow behind a section and a free traffic flow in front of this section neutralize the effect from forward propagation of heavy congestion, leading to less changes in velocities for this section. If, however, the section is affected by the backward propagation of heavy congestion, before the relieving of heavy congestion, the front congested traffic flow restricts this section to a small velocity without much possibility for vehicles to accelerate. This results in lower recovery rates for the velocity as well as in higher correlation of heavy congestion. The above phenomena are more clearly reflected in the ratio of the two rates $\beta_{ij,\mathrm{fit}}/\gamma_{ij,\mathrm{fit}}$, listed in table~\ref{tab4}. This ratio determines the dynamics of velocity changes. When $\beta_{ij,\mathrm{fit}}/\gamma_{ij,\mathrm{fit}}=1$, the decreased velocity due to heavy congestion and the increased velocity due to recovery balance out. In most cases, the ratio $\beta_{ij,\mathrm{fit}}/\gamma_{ij,\mathrm{fit}}>1$, suggesting the effect of heavy congestion prevails over the velocity recovery from the vehicle acceleration. The larger the ratio, the weaker the velocity adaptability on a section subject to the effect of heavy congestion. Evidently, the ratios are much higher in the case of backward propagation than in the case of forward propagation. As a result, the weak velocity adaptability on a section gives rise to the transient response standing out in the case of backward propagation of heavy congestion.

\begin{table}[tb]
\begin{minipage}{.49\linewidth}
 \centering
 \caption{The fitted propagation rates $\beta_{ij,\mathrm{fit}}$ }  
\begin{tabular}{c |ccccc}
\hlineB{2}
\diagbox{$i$}{$j$}& 1 & 2 & 3 & 4 & 5 \\
\hline
1   & \cellcolor{darkgray}{-}        &\cellcolor{lightgray}{0}    &\cellcolor{lightgray}{1.25}    &\cellcolor{lightgray}{1.38}    &\cellcolor{lightgray}{1.97}\\
 2  &1.19  &  \cellcolor{darkgray}{-}        &\cellcolor{lightgray}{0}    &\cellcolor{lightgray}{1.09}    &\cellcolor{lightgray}{0.84}\\
 3   &0.97   & 1.15        &  \cellcolor{darkgray}{-}   &\cellcolor{lightgray}{1.29}    &\cellcolor{lightgray}{0.79}\\
 4   &0.60    &1.24    &1.39        &  \cellcolor{darkgray}{-}        & \cellcolor{lightgray}{0}\\
 5   &0.13    &0.88    &1.76    &1.80        &  \cellcolor{darkgray}{-}\\
\hlineB{2}
\end{tabular}
\label{tab2} 
\end{minipage}
\begin{minipage}{.49\linewidth}
\centering
\caption{The fitted recovery rates $\gamma_{ij,\mathrm{fit}}$ }
\begin{tabular}{c|ccccc}
\hlineB{2}
\diagbox{$i$}{$j$}& 1 & 2 & 3 & 4 & 5 \\
\hline
1&\cellcolor{darkgray}{-}          &\cellcolor{lightgray}{0}    &\cellcolor{lightgray}{0.87}    &\cellcolor{lightgray}{1.00}    &\cellcolor{lightgray}{1.95} \\
 2&   0.16         &\cellcolor{darkgray}{-}     &\cellcolor{lightgray}{2.00}   &\cellcolor{lightgray}{0.72}    &\cellcolor{lightgray}{0.60}\\
 3&   0.19   & 0.30     &\cellcolor{darkgray}{-}     &\cellcolor{lightgray}{0.91}    &\cellcolor{lightgray}{0.54}\\
 4&   0.26    &0.42    &0.53         &\cellcolor{darkgray}{-}          &\cellcolor{lightgray}{0}\\
 5&  1.92    &0.56    &1.06    &0.99         &\cellcolor{darkgray}{-} \\
\hlineB{2}
\end{tabular}
 \label{tab3} 
\end{minipage}
\end{table}

\begin{table}[tb]
\caption{The ratio $\beta_{ij,\mathrm{fit}}/\gamma_{ij,\mathrm{fit}}$ }
\vspace*{-0.3cm}
\begin{center}
\begin{tabular}{c|ccccc}
\hlineB{2}
\diagbox{$i$}{$j$}& 1 & 2 & 3 & 4 & 5 \\
\hline
1&\cellcolor{darkgray}{-}      &\cellcolor{lightgray}{ -}   &\cellcolor{lightgray}{1.44}    &\cellcolor{lightgray}{1.38}    &\cellcolor{lightgray}{1.01}\\
2&7.44    &\cellcolor{darkgray}{-}    &\cellcolor{lightgray}{0}    &\cellcolor{lightgray}{1.51}    &\cellcolor{lightgray}{1.40}\\
3&5.11    &3.83    &\cellcolor{darkgray}{-}    &\cellcolor{lightgray}{1.42}   &\cellcolor{lightgray}{1.46}\\
4&2.31    &2.95    &2.62 &\cellcolor{darkgray}{-}      &\cellcolor{lightgray}{ -}\\
5&0.07    &1.57    &1.66  &1.82     & \cellcolor{darkgray}{-}\\
\hlineB{2}
\end{tabular}
\end{center}
\label{tab4}
\end{table}%

\section{Conclusions}
\label{sec5}

We introduced response functions as a new concept to study local dynamic in traffic networks, where the response function measures the velocity change of a section, on average, due to the heavy congestion of its neighbouring section. We found the dynamic characteristics of velocity responses depend on the propagation directions of heavy congestion. For the backward propagation, we found a transient response which manifests itself as a valley below zero values with the time lag. The transient response persists only a very short time period and performs unstably before a more stable long-term response sets in. For forward propagation, the transient response is almost absent and only the long-term response is present. 

The velocity transient response with the time lag is featured by valleys. Correspondingly, the congestion correlator with the time lag is featured by peaks. We found that the velocity response is inversely proportional to this congestion correlator, which might be viewed as a cause of the velocity response, regardless of other possible reasons. We also found the congestion correlator is mainly contributed by the heavy congestion with the duration around 100 min.

Inspired by the susceptible-infectious-recovered (SIR) model in epidemiology, we constructed a susceptible-decelerated-withdrawing (SDW) model for our case. The SDW model describes the dynamic characteristics of transient responses very well with the given initial values. Hence, the value of this model lies in information reduction, i.e. the model provides a set of filled parameters that fully characterize the transient response. The fitted parameters of propagation rates and recovery rates demonstrate that the heavy congestion on a section propagates forward and backward at a similar rate, but the forward sections are more likely to recover from the effect of heavy congestion than the backward sections. This is due to the weak velocity adaptability on backward sections, leading to the remarkable transient response in the case of backward propagation.  

For traffic management and planning as well as for the development of navigation systems, 
quantitative knowledge of traffic congestion, especially the heavy congestion, are indispensable. Response functions applied to traffic systems disclose the velocity change of one road section based on the prior traffic congestion occurring at a different road section. Such information known in advance can be a precursor of traffic phases of a given road section. This might be helpful for traffic management or navigation systems with congestion pre-warning. Either for human or autonomous driving, to avoid or prevent congestions or enhance the traffic efficiency, prior information without much complex computation for designing an optimal driving route will be an advantage in the future. Response functions applied to traffic systems appear as a promising tool in this context.

\section*{Acknowledgements}

We gratefully acknowledge funding via the grant ``Korrelationen und deren Dynamik in Autobahnnetzen'', Deutsche Forschungsgemeinschaft (DFG, 418382724). We thank Strassen.NRW for providing the empirical traffic data. We also thank Henrik M. Bette and Sebastian Gartzke for fruitful discussions.

\section*{Author contributions}

T.G. and M.S. proposed the research. S.W. and T.G. developed the methods of analysis. S.W. performed all the calculations. S.W. and T.G. wrote the manuscript with the input from M.S. All authors contributed equally to analyzing the results and reviewing the paper.


\end{document}